\documentclass[referee]{aa} 

\usepackage{graphicx}
\usepackage{url}
\usepackage{hyperref}
\hypersetup{
    colorlinks=true,
    linkcolor=red,
    filecolor=magenta,      
    urlcolor=magenta,
    citecolor=blue,
}
\usepackage{txfonts}
\usepackage{rotating}
\usepackage{lscape}
\usepackage{multirow}
\begin{document} 

   \title{Type II radio bursts and their association with coronal mass ejections in solar cycles 23 and 24}

   \titlerunning{Type II and CMEs}

   \author{A. Kumari  \inst{1,2}
   \and
    D. ~E.~Morosan \inst{1} 
    \and
    E.~K.~J.~Kilpua \inst{1} 
    \and
    F. Daei \inst{1}
    }

   \institute{Department of Physics, University of Helsinki, P.O. Box 64, FI-00014, Helsinki, Finland \\
      \and
   NASA Postdoctoral Program Fellow, NASA Goddard Space Flight Center, Greenbelt, MD 20771, USA \\
              \email{anshu.kumari@helsinki.fi}
          }

   \date{Received ; accepted }


 
  \abstract
   {Metre wavelength type II solar radio bursts are believed to be the signatures of shock-accelerated electrons in the corona. Studying these bursts can give information about the initial kinematics, dynamics and energetics of CMEs in the absence of white-light observations. 
   }
   {In this study, we investigate the occurrence of type II bursts in solar cycles 23 and 24 and their association with coronal mass ejections (CMEs). We also explore the possibility of occurrence of type II bursts in the absence of a CME. 
   }
   {We performed statistical analysis of type II bursts that occurred between 200 - 25 MHz in solar cycle 23 and 24 and found the temporal association of these radio bursts with CMEs. We categorised the CMEs based on their linear speed and angular width, and studied the distribution of type II bursts with `fast' ($speed ~\geq 500 km/s$), `slow' ($speed ~< 500 km/s$), `wide' ($width ~\geq 60^o$) and `narrow' ($width ~< 60^o$)  CMEs. We explored the type II bursts occurrence dependency with solar cycle phases. 
   }
   {Our analysis shows that there were total 768 and 435 type II bursts in solar cycle 23 and 24, respectively. Of these, 79 $\%$ were associated with CMEs in solar cycle 23 and 95\% were associated with CMEs in solar cycle 24. However, only 4$\%$ and 3$\%$ of the total number of CMEs were accompanied by type II bursts in solar cycle 23 and 24, respectively. Most of the type II bursts in both cycles were related to `fast and wide' CMEs (48 $\%$). We also determined a typical drift rate and duration for type II bursts, which is 0.06 MHz/s and 9 min, respectively. Our results suggest that type II bursts dominate at heights $\approx 1.7 - 2.3 \pm 0.3 ~R_{\odot}$ with a clear majority having an onset height around 1.7 $\pm 0.3~R_{\odot}$ assuming the four-fold Newkirk model.
   }
   {The results indicate that most of the type II bursts had a white-light CME counterpart, however there were a few type II which did not have a clear CME association. There were more CMEs in cycle 24 than cycle 24. However, the number of type II radio bursts were less in cycle 24 compared to cycle 23. The onset heights of type IIs and their association with wide CMEs reported in this study indicate that the early CME lateral expansion may play a key role in the generation of these radio bursts.
   
   }


   \maketitle
%

\section{Introduction}
\label{section:sec1}

Coronal mass ejections (CMEs) can be powerful drivers of plasma shocks that can accelerate electrons to high energies. Accelerated electron beams can in turn generate radio bursts at metre and decimetre wavelengths through the plasma emission mechanism. Radio shock signatures associated with CMEs are categorised as type II radio bursts \citep{Smerd1970, Dulk1970, kl02, Gopalswamy2005, Ramesh2023a}.
Type II radio bursts are characterised as lanes of emission drifting from high to low frequencies in the solar dynamic spectra, with a duration of a few minutes to approximately one hour. These lanes have a frequency ratio of 1 to 2 representing emission at the fundamental and harmonic of the plasma frequencies and \citep{ne85}. Radio imaging and spectroscopy combined with coronal modelling have shown that type II bursts are associated to CME-driven shock waves \citep[e.g.,][]{Smerd1970, Ramesh2010a, Zucca2018, kumari2019direct}. Type II bursts can also consist of numerous fine structures. For example, herringbone bursts are narrow bursts in dynamic spectra superimposed on type II bursts or occurring on their own, drifting towards higher and lower frequencies \citep{Roberts1959,ca87,Carley2015}. Herringbones are believed to be the radio signature of individual electron beams accelerated by the CME shock \citep{Zlobec1993, Morosan2019a}. A wide variety of other fine structures composing the emission lanes of type II bursts have recently been documented in \citet{Magdalenic2020}. 

\begin{figure*}[ht!]
   \centering
  \includegraphics[width=1\textwidth,clip=]{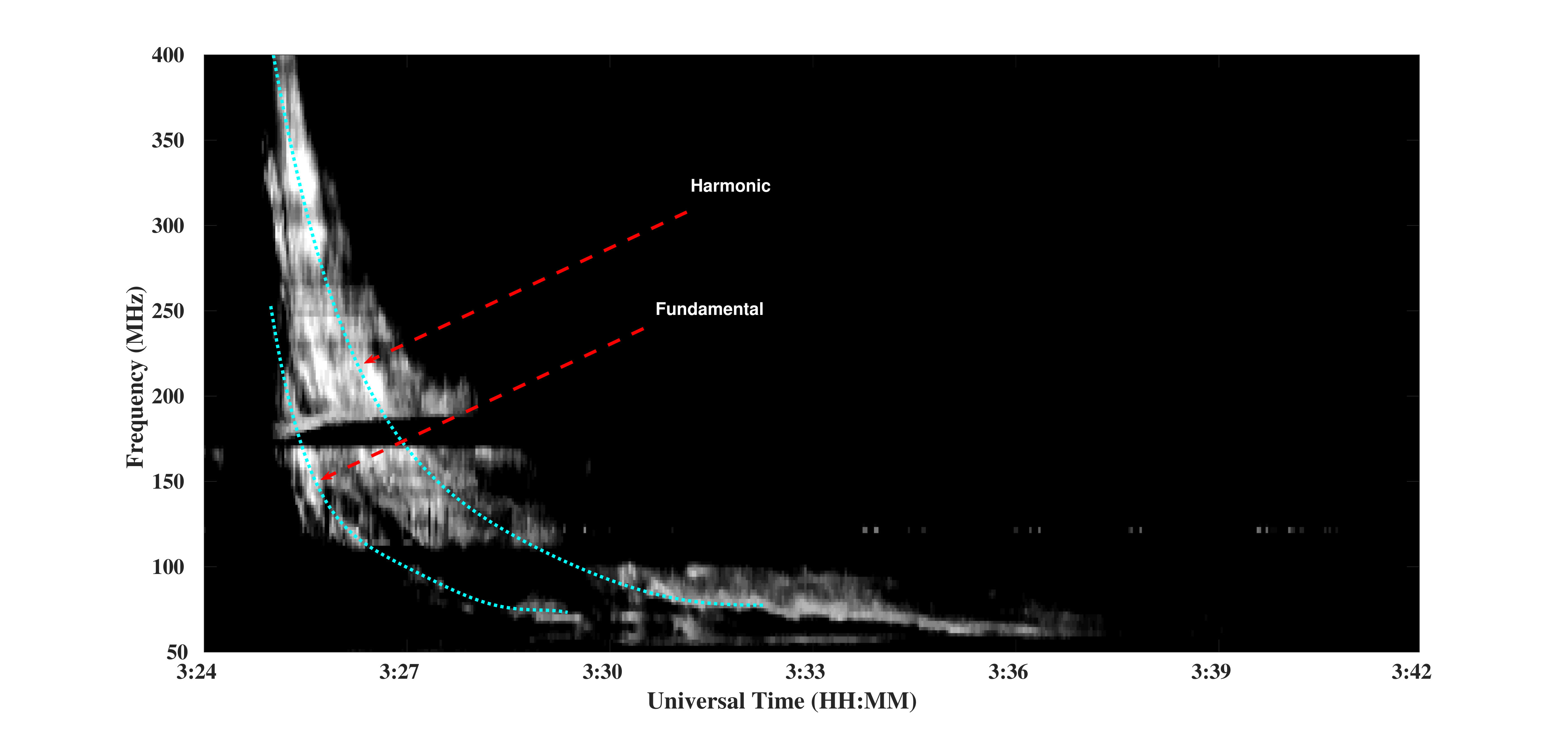}
      \caption{Solar radio dynamic spectra of a high frequency type II burst on November 04, 2015. The duration of the burst is $\sim 12 $ minutes and it shows a frequency drift of $\sim 0.2$ MHz/s from 400 MHz to 55 MHz. The `cyan' color lines show the fundamental and harmonic bands in the spectra. 
      }
         \label{Fig:figure1}
\end{figure*}

Type II bursts are the most common radio bursts accompanying CMEs, hence these bursts are often used to study CME kinematics, energetics and dynamics near the Sun \citep{Cunha-Silva2015, kumari2017b, Ma2020}. The advancement of space-based coronagraphs and ground-based spectrographs in the past two decades helped to understand the relationship between CMEs and meter wavelength type II radio bursts. Metric type II bursts start typically at $\approx 1.5~ R_{\odot}$ \citep{lin2006theoretical}.
\cite{Clasen2002} studied the association of metric and interplanetary (IP) type II bursts with CMEs observed with LASCO for 63 events during 1997-2000, where the authors related the type II origin to blast wave shocks or the leading edge/flanks of the CMEs. \cite{Gopalswamy2006} performed a similar study for metric and  decameter-hectometer (DH) type II bursts during 1996-2004, where the author concluded that the metric type II bursts are mostly associated with slow CMEs and DH type II are associated with CMEs with speed $> 1100$ km/s. \cite{Kahler2019} also studied metric and DH type II bursts and their association with solar energetic particles (SEPs) and found that it is uncommon for these events to be associated with `fast' and `narrow' CMEs. 
Recently, \cite{Pohjolainen2021} studied isolated DH type II radio bursts, which were found to be associated with a shock near the CME leading front.

However, at present there is no comprehensive long term statistical study of metric type II bursts and their relationship with CMEs that covers over one solar cycle. 
Recently, \cite{Morosan2021} studied moving radio bursts, that include type II and moving type IV bursts, and their association with CMEs, and found that moving radio sources are almost exclusively associated with CME eruptions. \cite{Patel2021} studied the decameter-hectometer type II bursts and concluded that CME initial speed are related to the duration of type  II  bursts. 
With the end of solar cycle 24 and the availability of white-light CME and type II radio bursts data for more than two decades, we take this opportunity to study the association of type II bursts with CMEs for the last two solar cycles, 23 and 24, and determine general properties of these CMEs and the associated radio emission. This article is arranged as follows: section \ref{section:sec2} has the details of observational data and the data analysis methods. We discuss the results obtained with our analysis in section \ref{section:sec3}. We present the summary and conclude the paper in the last section \ref{section:sec4}.

   \begin{figure*}[ht]
    \centering \includegraphics[width=1\textwidth,clip=]{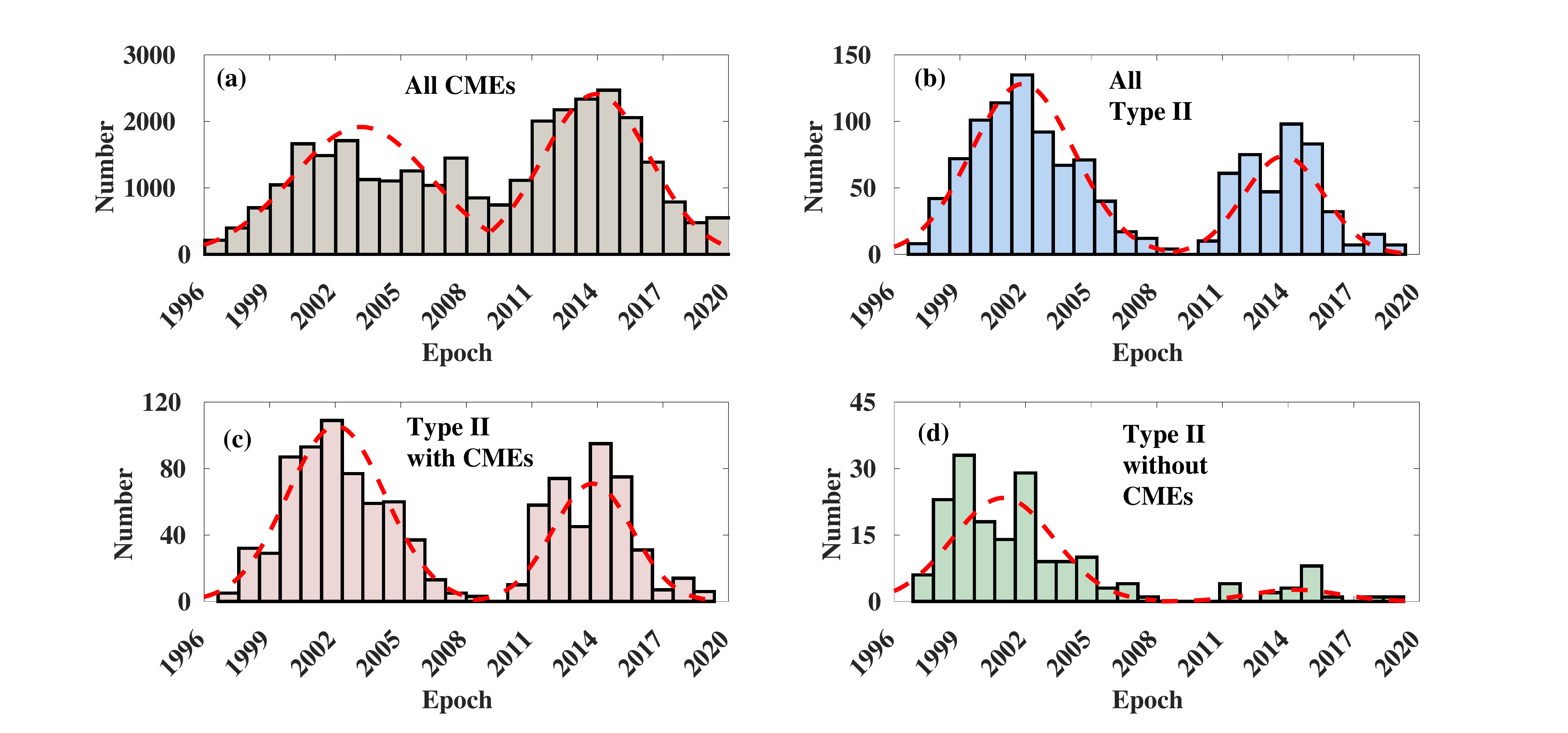}
   \caption{ The histograms showing the distribution of occurrence of: (a) all CMEs; (b) all type II bursts; (c) all type II bursts which were accompanied with white-light CMEs; and d) all type II bursts without any association with CME;
   from January, 1996 to December, 2019, which corresponds to solar cycle 23 and 24.
   The dashed red profile shows the Gaussian fit to the the histograms. The two peaks of these Gaussian peaks correspond to the solar maxima in the respective solar cycles. }
    \label{Fig:figure2}
    \end{figure*}

\section{Observations and Data Analysis}
\label{section:sec2}

We used the CMEs listed in the Coordinated Data Analysis Workshop \citep[CDAW;][]{Yashiro2004,Yashiro2008,Gopalswamy2009cdaw} database\footnote{\url{https://cdaw.gsfc.nasa.gov/}}, which lists the CMEs manually detected with the the Large Angle and Spectrometric Coronagraph (LASCO) onboard the Solar and Heliospheric Observatory \citep[SOHO;][]{Brueckner1995}. This list contains the CMEs detected with LASCO coronagraphs since January, 1996. It includes CME properties such as linear and second order speed, position angle, acceleration, mass, and kinetic energy. We also used the Cor1 and Cor2 coronagraphs from the Sun-Earth Connection Coronal and Heliospheric Investigation (SECCHI; \citet{Howard2008}) onboard the Solar Terrestrial Relationship Observatory (STEREO). For the CMEs observed with the STEREO spacecrafts, we used the list provided on the Solar Eruptive Event Detection System (SEEDS)\footnote{\url{http://spaceweather.gmu.edu/seeds/secchi/detection_cor2/monthly/}} website. The CMEs listed here are detected automatically using software packages, which extracts the position angle, angular width, velocity, peak, and average brightness \citep{Olmedo2008}. We also used CME list provided by \citet{Vourlidas2017}, where the authors used the  dual-viewpoint CME catalog from the SECCHI/COR telescopes\footnote{\url{http://solar.jhuapl.edu/Data-Products/COR-CME-Catalog.php}}. These catalogues contain STEREO detected CMEs from 2009 - 2014. 

For radio data, we used the event lists provided by the Space Weather Prediction Center (SWPC)\footnote{\url{https://www.swpc.noaa.gov/products/solar-and-geophysical-event-reports}}. These event lists contain details of the solar eruptions such as X-ray and optical flares, radio burst types and their proprieties, the start and end time of the eruption, the class of the flares, frequency of radio bursts etc. We also used the sunspot data to investigate the correlation between the appearance of sunspots and the occurrence of type II bursts. The sunspot data was taken from \cite{sidc}\footnote{\url{http://www.sidc.be/silso/datafiles}}, which has daily, monthly mean and 13 month averaged sunspots data available. 

In this work, we investigate the relation between type II bursts and CMEs in solar cycle 23 and 24. To do this, we extracted the type II bursts information from the SWPC event lists from January, 1996 - December, 2019. This exclusive type II burst list contains all the type II bursts in solar cycle 23 and 24, with details such as, start and end time, start and end frequency, reporting spectrograph, and the associated active region (AR). 
An example of a type II burst is shown in Figure \ref{Fig:figure1}a. This figure shows a high frequency, high drift type II burst that occurred on November 04, 2015. The solar dynamic spectra shown here is with the ground-based Compound Astronomical Low-cost Low-frequency Instrument for Spectroscopy and Transportable Observatory (CALLISTO)\footnote{\url{http://www.e-callisto.org/}} spectrometer at the Gauribidanur Radio Observatory (GRO)\footnote{\url{https://www.iiap.res.in/centers/radio}}.
The fundamental and harmonic bands are the type II bursts and they are marked in the figure. 

To find the type II associations with CMEs, we used the criteria that the start time of the associated CME should be within 2 hours of the start time of the type II. This time difference between the start times of type IIs and CMEs was chosen in order to compensate for disc events where the CME may be seen later in white light images compared to limb events \citep[e.g.,][]{Kumari2021, Morosan2021, Kumari2022}. 
Two new list was prepared  based on the aforementioned criteria, which contain  the information of: i) the type II bursts associated with CMEs; ii) the type II bursts without any association with CMEs. These two lists were used to further carry out the current study. 

\begin{table*}[]
\caption{Yearwise distribution of CMEs, type II radio bursts and their association.}
\label{tab:table1}
\begin{tabular}{ccccccccc}
\hline
Year                 & CMEs                 & \% CMEs              & Type IIs             & \% Type IIs          & Type IIs                      & \% Type IIs                   & Type IIs                        & \% Type IIs                     \\
\multicolumn{1}{l}{} & \multicolumn{1}{l}{} & \multicolumn{1}{l}{} & \multicolumn{1}{l}{} & \multicolumn{1}{l}{} & \multicolumn{1}{l}{with CMEs} & \multicolumn{1}{l}{with CMEs} & \multicolumn{1}{l}{without CME} & \multicolumn{1}{l}{without CME} \\
\hline
Total        & 30125  & 100.0\%   &    1203     & 100.0\%       & 1024                 & 85.2\%                 & 179                     & 14.8\%  \\
\hline
1996         & 206   & 1.5\%   & 4        & 0.5\%       & 2                  & 0.3\%                 & 2                     & 1.2\%                    \\
1997         & 385   & 2.7\%   & 21       & 2.7\%       & 16                 & 2.6\%                 & 5                     & 3.1\%                    \\
1998         & 716   & 5.1\%   & 60       & 7.8\%       & 24                 & 3.9\%                 & 36                    & 22.6\%                   \\
1999         & 1016  & 7.2\%   & 100      & 13.0\%      & 75                 & 12.3\%                & 25                    & 15.7\%                   \\
2000         & 1664  & 11.9\%  & 90       & 11.7\%      & 72                 & 11.8\%                & 18                    & 11.3\%                   \\
2001         & 1499  & 10.7\%  & 160      & 20.8\%      & 136                & 22.3\%                & 24                    & 15.1\%                   \\
2002         & 1700  & 12.1\%  & 116      & 15.1\%      & 96                 & 15.8\%                & 20                    & 12.6\%                   \\
2003         & 1130  & 8.1\%   & 71       & 9.2\%       & 62                 & 10.2\%                & 9                     & 5.7\%                    \\
2004         & 1102  & 7.9\%   & 69       & 9.0\%       & 58                 & 9.5\%                 & 11                    & 6.9\%                    \\
2005         & 1248  & 8.9\%   & 49       & 6.4\%       & 45                 & 7.4\%                 & 4                     & 2.5\%                    \\
2006         & 1047  & 7.5\%   & 19       & 2.5\%       & 15                 & 2.5\%                 & 4                     & 2.5\%                    \\
2007         & 1442  & 10.3\%  & 7        & 0.9\%       & 6                  & 1.0\%                 & 1                     & 0.6\%                    \\
2008         & 863   & 6.1\%   & 2        & 0.3\%       & 2                  & 0.3\%                 & 0                     & 0.0\%                    \\
\hline
Total (SC23) & 14018 & 100.0\% & 768      & 100.0\%     & 609                & 79.3\%               & 159                   & 20.7\%                  \\
\hline
2009         & 746   & 4.6\%   & 1        & 0.2\%       & 1                  & 0.0\%                 & 0                     & 0.0\%                    \\
2010         & 1117  & 6.9\%   & 13       & 3.0\%       & 13                 & 3.1\%                 & 0                     & 0.0\%                    \\
2011         & 1990  & 12.4\%  & 72       & 16.5\%      & 68                 & 16.4\%                & 4                     & 20.0\%                   \\
2012         & 2177  & 13.5\%  & 71       & 16.3\%      & 69                 & 16.6\%                & 2                     & 10.0\%                   \\
2013         & 2338  & 14.5\%  & 73       & 16.9\%      & 72                 & 17.3\%                & 1                     & 5.0\%                    \\
2014         & 2478  & 15.4\%  & 110      & 25.3\%      & 102                & 24.6\%                & 8                     & 40.0\%                   \\
2015         & 2058  & 12.8\%  & 54       & 12.4\%      & 51                 & 12.9\%                & 3                     & 15.0\%                   \\
2016         & 1393  & 8.7\%   & 15       & 3.4\%       & 15                 & 3.6\%                 & 0                     & 0.0\%                    \\
2017         & 786   & 4.9     & 18       & 4.1\%       & 6                  & 4.1\%                 & 1                     & 5.0\%                    \\
2018         & 476   & 2.9\%   & 2        & 0.5\%       & 2                  & 0.5\%                 & 0                     & 0.0\%                    \\
2019         & 543   & 3.4\%   & 6        & 1.4\%       & 1                  & 1.2\%                 & 1                     & 5.0\%                    \\
\hline
Total (SC24) & 16107 & 100.0\% & 435      & 100.0\%     & 415                & 95.4\%               & 20                    & 4.6\%    \\
\hline

\end{tabular}
\end{table*}

\section{Results}
\label{section:sec3}

 The yearly distribution of CMEs and type II radio bursts, and the percentage of type II bursts with and without CMEs are shown in Table \ref{tab:table1}. There were 768 and 435 type II bursts reported during SC 23 and SC 24, respectively. There were 14018 and 16107 CMEs reported during SC 23 and SC 24, respectively (based on LASCO white-light observations). There were additional 5944 CMEs recorded with the STEREO coronagrapghs (A and B combined) in SC 24.

The results in Table \ref{tab:table1} are also illustrated in Figure \ref{Fig:figure2} which shows the histograms of occurrence of CMEs and type II bursts in cycles 23 and 24. We fit a Gaussian to these distributions which showed that both CMEs and type II events peaks during solar cycle maxima, i.e during the years 2000-2002 and 2013-2014 for cycle 23 and 24, respectively. We note that for cycle 23, there were more CMEs during the declining phase than rising phase of cycle. Almost $\approx 24 \%$ of the CMEs in cycle 23 were recorded during the decline phase. In contradiction, there were more CMEs during the rising phase of cycle 24 than the declining phase. However, for type II radio bursts, both cycles showed similar trend, where there were more bursts during the beginning of the cycle compared to the end of the cycle. 
This indicates that the type II bursts occurrence follow the same pattern as sunspot cycle.


\subsection{Type II Bursts and associated CMEs} 
\label{section:sec3.1}

In cycle 23, $\approx 79\%$ type II bursts were associated with a white-light CME using our criteria, while in cycle 24 the percentage was considerable higher,  $\approx 95\%$s. However, the fraction of CMEs associated with type II bursts were more similar and very low, only $\approx 4\%$ and $\approx 3\%$ in SC23 and SC 24, respectively. The lower association of type II bursts to CMEs in SC 23 could be explained by two possibilities: i) the unavailability of the STEREO spacecraft that were available during SC 24 to provide different view points making it possible to identify more CMEs especially those originating from the far-side of the Sun; and ii) the loss of contact of LASCO during 1998 when no white-light CME identifications could be made. We note that the loss of contact with STEREO-B during SC4, we could not use it for type II and CME association.
Figure \ref{Fig:figure2} (bottom panel) shows the distribution of occurrence of type II bursts with and without any CME association. The Gaussian fits show a similar pattern where the peaks coincide with then peak of the solar cycle. Table \ref{tab:table1} also shows the type II-CME association for both the cycles.

   \begin{figure*}
    \centering \includegraphics[width=1\textwidth,clip=]{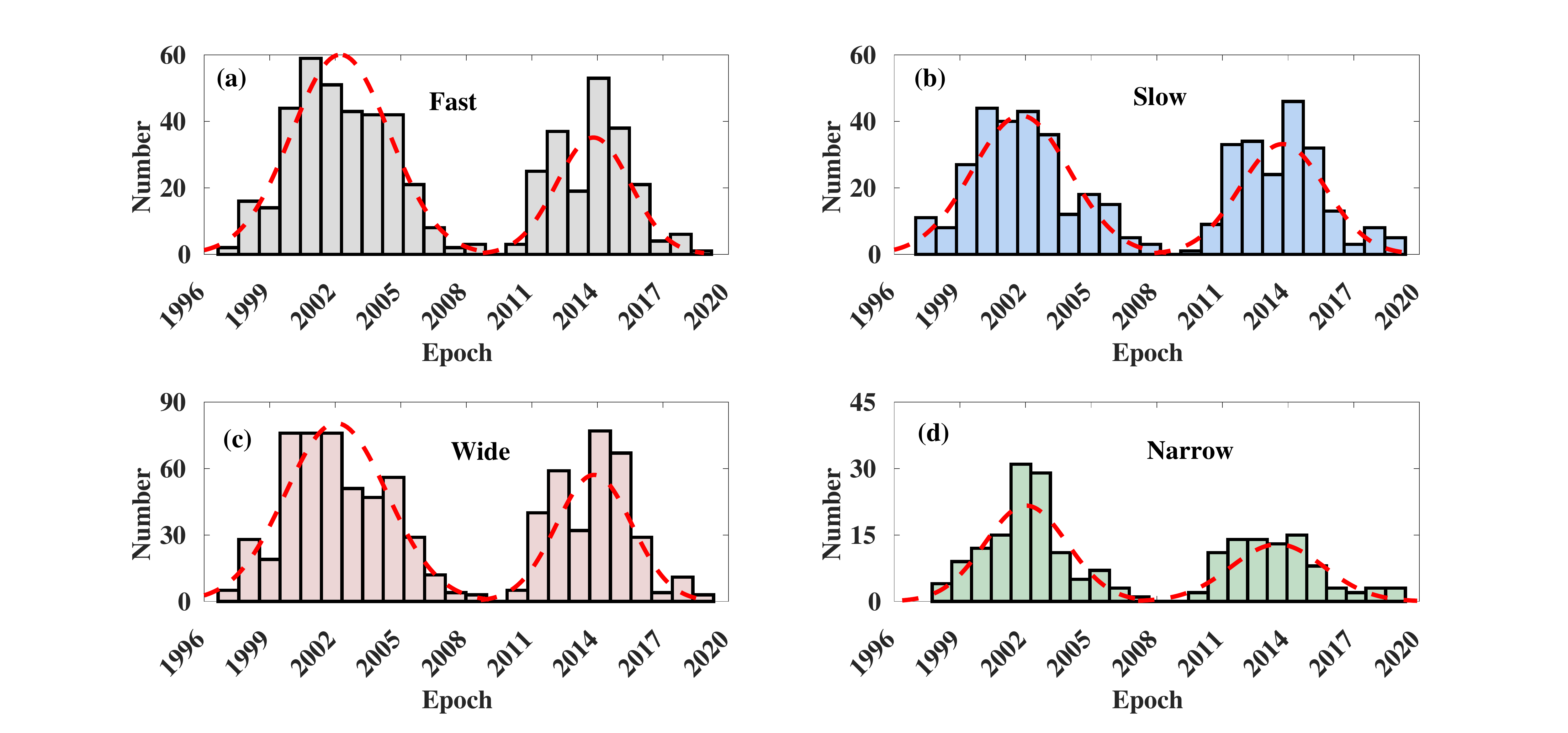}
    \centering \includegraphics[width=1\textwidth,clip=]{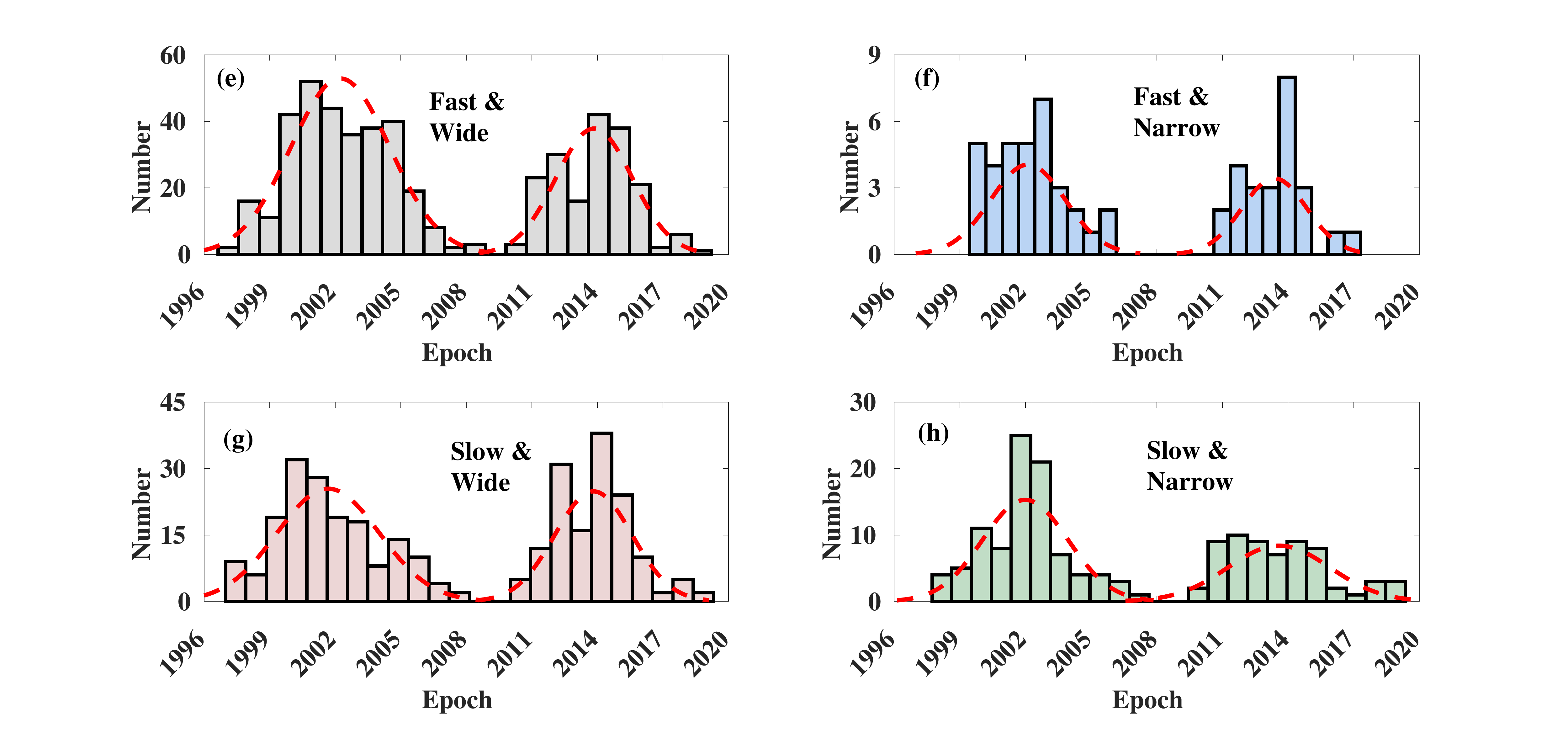}
   \caption{ The histograms showing the variation of occurrence of type II bursts with: (a) Fast; (b) Slow; (c) Wide; (d) Narrow; (e) Fast and Wide; (f) Fast and narrow; (g) Slow and Wide; (h) Slow and Narrow CMEs for solar cycle 23 and 24.}
    \label{Fig:figure3}
    \end{figure*}

\subsection{Type II Bursts associated with type of CMEs}
\label{section:sec3.2}
We categorised the CMEs on the basis of their linear speeds and angular width as `Fast', `Slow', `Wide' and `Narrow' CMEs \citep{Kumari2021, Morosan2021}. The CMEs which had linear speed $\geq 500$ km/s were classified as `Fast' CMEs, and the rest were kept in the 'Slow' CME category. Similarly, CMEs with angular width $\geq 60^o$ were designated as `Wide' CMEs and the rest were `Narrow' CMEs. Table \ref{tab:table2} contains the list of type II bursts associated with different type of CMEs.  

We found that most of the type II bursts were associated with 'Fast and Wide' CMEs, irrespective of the cycle ($\approx 48 \%$).  Figure \ref{Fig:figure3} shows histograms of the
distribution of type II bursts associated with various CME
types.  The Gaussian distributions to the dataset peak during solar maximum for 'Fast', 'Slow', 'Wide', and 'Narrow' CMEs for both solar cycles. We note that 'Fast' and 'Slow' CMEs were almost equally distributed in the past two cycles, however most of the type II bursts were associated with 'Wide' CMEs. 
The bottom panel of Figure \ref{Fig:figure3} also shows the type II bursts for a combination of speed and width of CMEs.
There were a very few type II bursts associated with 'Fast and Narrow' CMEs. Figure \ref{Fig:figure3} and Table \ref{tab:table2} show that type II bursts association with CMEs follows the same pattern in both solar cycles. 

We note that we have considered the linear speed of CMEs as mentioned in the CME catalogues used in the present study. A constant linear speed has projection effects and it does not include the complete kinematics of CMEs in the solar atmosphere. However, we used an average linear speed for CMEs since we have more than thirty thousand CMEs included in our study. We acknowledge that this could lead into a few misclassification for type of CMEs in the present study. Several previous studies have reported that type II bursts are often related to the flanks of CMEs \citep{Holman1983, Gopalswamy2009, kumari2017b}. The CME speeds at the leading edge and flanks may be vary and individual case studies needs to be carried out to verify the CME shock speed and type II association.

           \begin{figure*}
    \centering \includegraphics[width=1\textwidth,clip=]{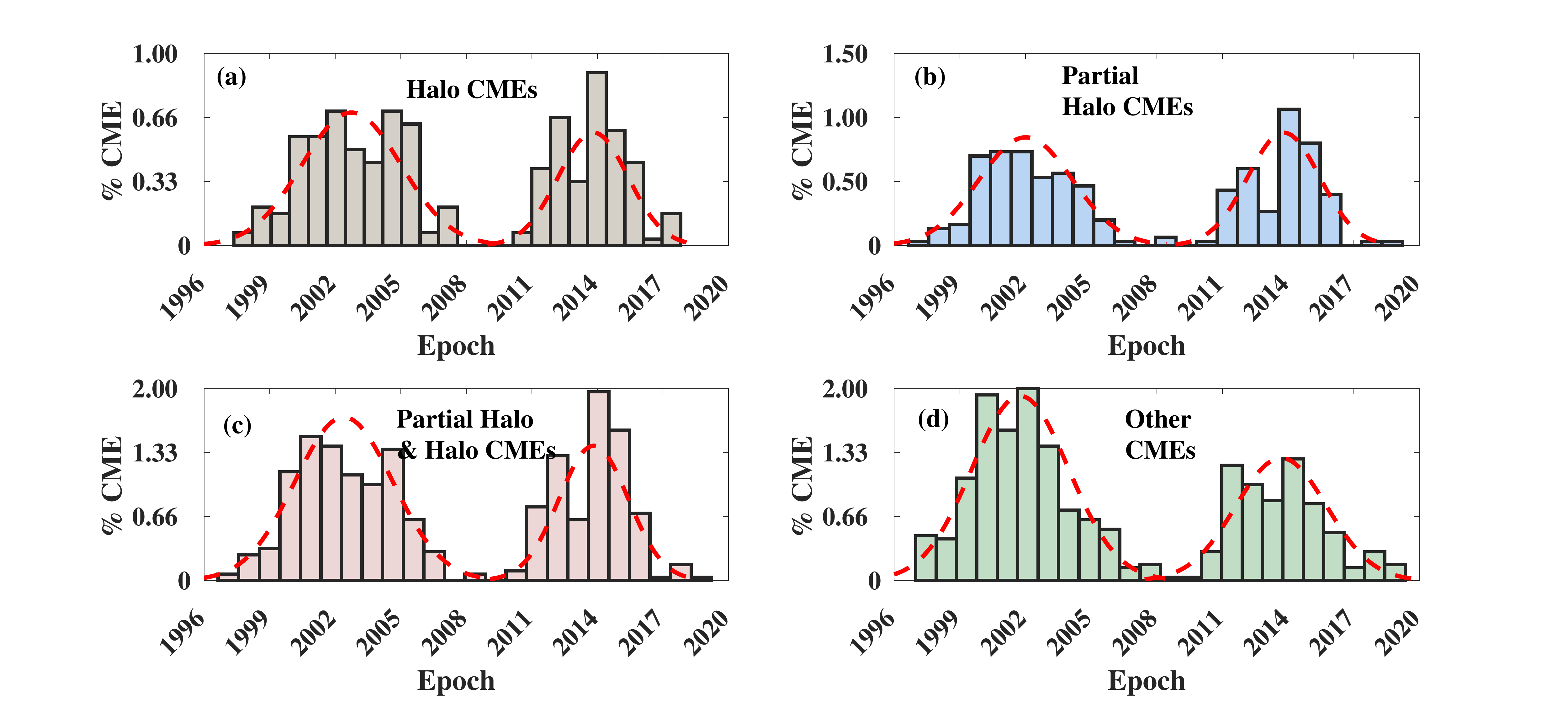}
   \caption{ The histograms showing the distribution of occurrence type II bursts with: (a) all Halo CMEs; (b) all partial Halo CMEs; (c) all partial Halo and Halo CMEs; and d) all other CMEs;
   from January, 1996 to December, 2019, which corresponds to solar cycle 23 and 24.
   The dashed red profile shows the Gaussian fit to the the histograms. The two peaks of these Gaussian peaks correspond to the solar maxima in the respective solar cycles. }
    \label{Fig:figure4}
    \end{figure*}

\subsection{Type II Bursts associated with Halo and Partial Halo CMEs}
\label{section:sec3.3}

We studied the association of type II radio bursts with halo CMEs. partial halo and halo CMEs are included into the 'Wide' category. The CMEs with width $W= 360^o$ are classified as full halos and those with width $120^o \leq W < 360^o$ are classified as partial halos. 
Table \ref{tab:table3} lists the type II bursts associated with partial and full halo CMEs in cycle 23 and 24.
We found that there were more type II bursts associated with partial and full halo CMEs in cycle 24 $26 \%$) than cycle 23, irrespective of the fact that there were less type IIs in the latter cycle. This 
The distribution halo CMEs were almost equally distributed as partial and full halos for both the cycle. 
Figure \ref{Fig:figure4} shows the distribution of type II bursts associated with partial halo, full halo and other CMEs. The Gaussian fits to the distribution also follow similar pattern as previous distribution i.e the peaks occur during the maxima of the respective solar cycles. 

       \begin{figure*}
    \centering \includegraphics[width=1\textwidth,clip=]{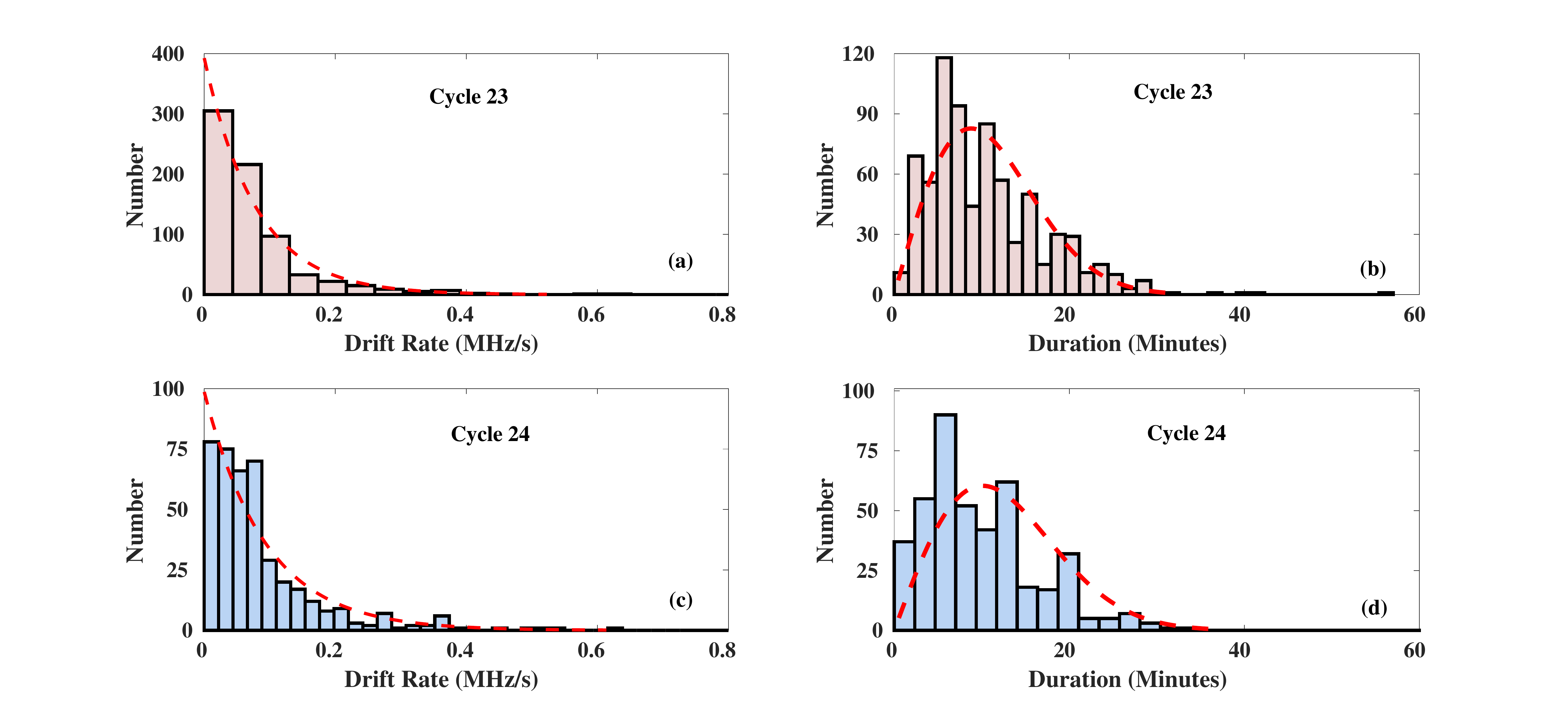}
   \caption{The histograms showing the variation drift rates and duration of type II bursts solar cycle 23 ((a) and (b), respectively) and 24 ((c) and (d), respectively).}
    \label{Fig:figure5}
    \end{figure*}

           \begin{figure*}
    \centering \includegraphics[width=1\textwidth,clip=]{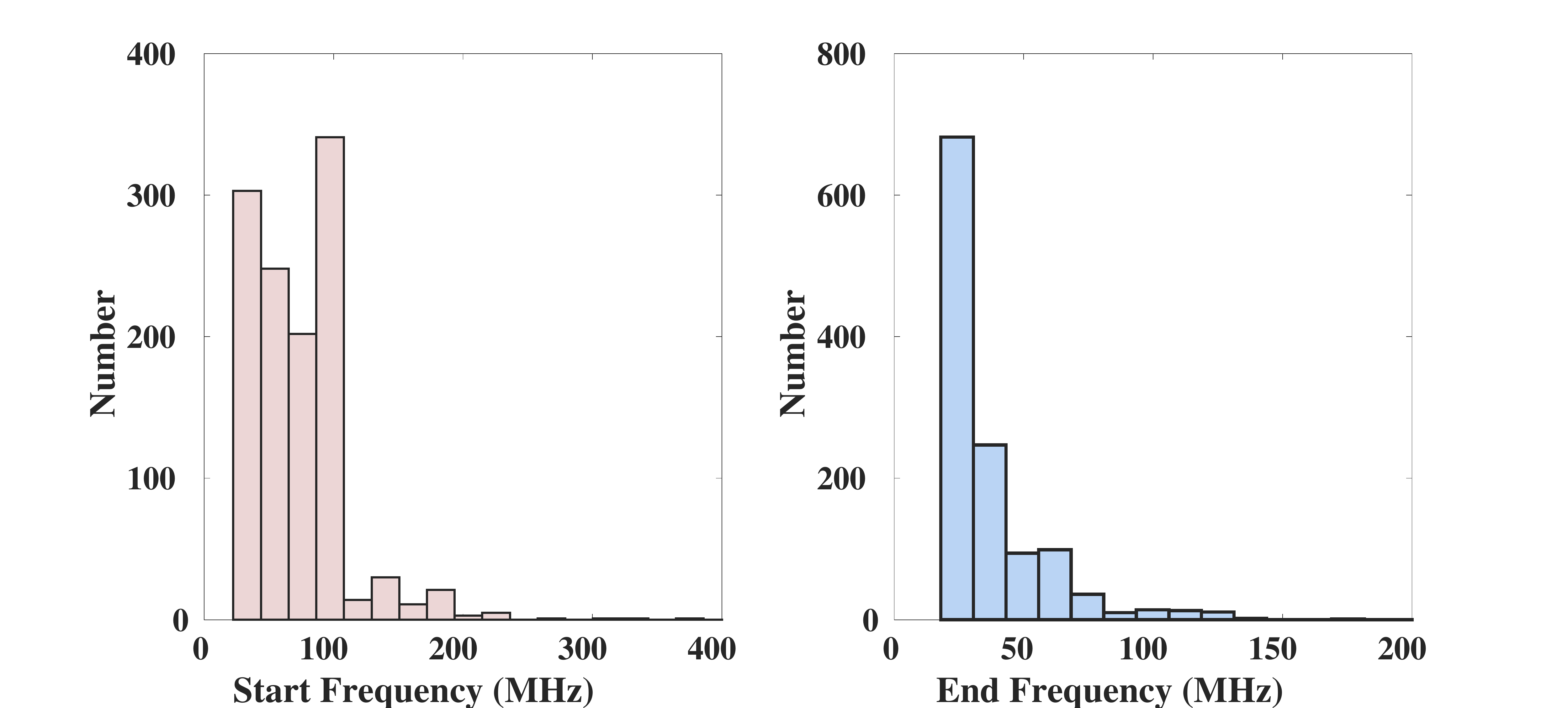}
   \caption{The histograms showing the variation of the start (left panel) and end (right panel) frequencies of type II bursts in cycle 23 and 24. We have used the fundamental start frequency of all the type II bursts as mentioned in section \ref{section:sec3.5}. For the type II bursts which were a fundamental-harmonic pair, we have divided the start frequencies by a factor of two. }
    \label{Fig:figure5a}
    \end{figure*}

\subsection{Type II Bursts and Their Properties}
\label{section:sec3.4}

We analysed the spectral (time-frequency) properties of the type II bursts for solar cycle 23 and 24 using the type II bursts list generated in section \ref{section:sec2}.
We estimated the duration of the type II bursts using the start and end time ($t_{start}$ and $t_{end}$, respectively)  as noted in the SWPC event lists. We also estimated the drift rates of these bursts using the start and end frequencies ($f_{start}$ and $f_{start}$, respectively) as recorded in the solar dynamic spectra. 
Figure \ref{Fig:figure5} shows the distribution of drift rate and duration of the type II bursts in the last two cycles. More than $90 \%$ of the type II bursts had drift rates higher than 0.05 MHz/s in both cycles for frequency range 200 - 25 MHz. The behaviour of these type II bursts are very similar to moving type IV bursts as reported by \cite{Kumari2021}.
Similarly, almost $80 \%$ bursts had duration less than 20 minutes. We note that type II bursts are slowing drifting features on the solar dynamic spectra, and these can have different drift rates, in turn different speeds in the solar atmosphere. In the present study, we have considered a constant drift rate for type II bursts, which can underestimate or overestimate the shock speed calculated from the radio burst. 

       \begin{figure*}
    \centering \includegraphics[width=1\textwidth,clip=]{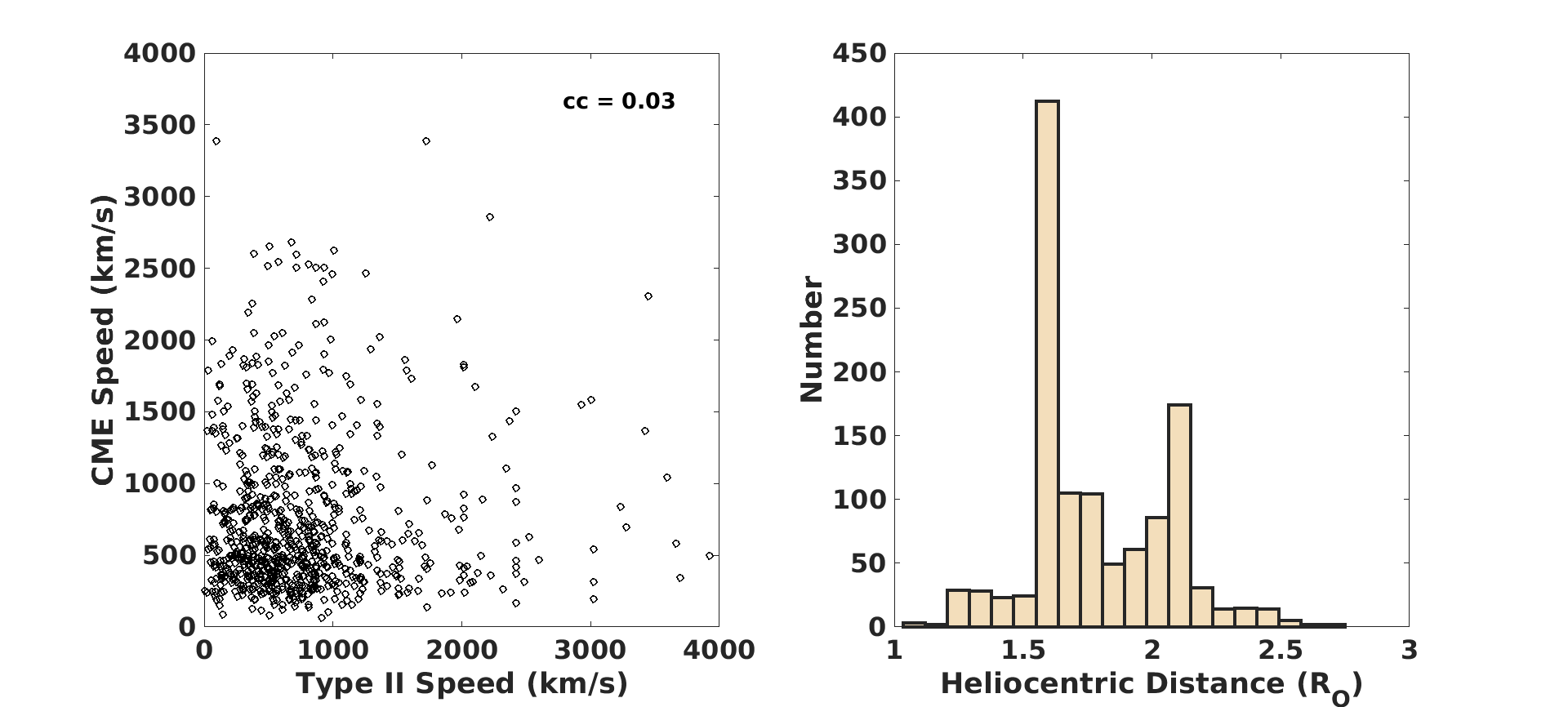}
   \caption{ \textbf{Left panel:} The relation between CME linear speeds as mentioned in catalogue and the type II speed estimated with four-fold Newkirk's density model and duration of the bursts.  
   \textbf{Right panel:} Distribution of type II formation heights using the aforementioned density model.
   }
    \label{Fig:figure6}
    \end{figure*}

\subsection{Heights of Type II bursts}
\label{section:sec3.5}
We converted the start and end frequencies obtained for the type II busts from the SWPC events into heliocentric distances. We used the four-fold Newkirk's density model for frequency to heliocentric distance conversion \citep{newkirk1961solar, Mann1995}. 
We used this model and enhancement factor to account for the enhanced densities above active regions. There is always an ambiguity in using a particular density model as different density models and enhancement factors can reflect a range of possible heights of the radio sources. 
We began with an assumption that all the type II bursts were a fundamental-harmonic pair, hence the start frequency of the burst was divided by a factor of two for a more accurate height estimation. The end frequency of the bursts was kept as it is since in the end frequency of type II bursts is the end frequency of the fundamental lane. The speeds of the type II bursts were estimated using the height estimate and the time duration (see section \ref{section:sec3.4}) of the bursts. 
Figure \ref{Fig:figure5a} shows the histogram of start and end frequencies of all the type II bursts in studied in this manuscript. For the type II bursts which were a fundamental-harmonic pair, we have divided the start frequencies by a factor two. 

We first compared the speed of the type II bursts to that of the associated CME. Figure \ref{Fig:figure6}a shows the linear correlation between the shock speed from radio bursts and CME speed from whitelight coronagraph data. There was almost no correlation (cc =0.03) found between type II speed and CME linear speed. The most likely cause for this lack of correlation is that the speed of the metric type II bursts reflects the speed at  lower coronal heights ($\leq 2.5 ~R_{\odot}$) compared to the CMEs linear speed mentioned in the CDAW catalogue which is obtained at much higher heights ($\geq 2.5 ~R_{\odot}$). The type II speeds are also likely to be indicative of the lateral or flank speeds of CMEs and not the radial outward speed since a lot of metric type IIs have been reported to occur at the CME flank \citep{Chrysaphi2018,Morosan2019a,Majumdar2021}. Another likely cause of low correlation could be due to the assumption of constant plane-of-sky CME speed.
Figure \ref{Fig:figure6}b shows the distribution of type II heights over the years. We found that by using four-fold Newkirk's density model, for a large majority of type IIs, their starting height is at $\approx 1.7 \pm 0.3 ~R_{\odot}$ as shown by the significant peak in \ref{Fig:figure6}b (for frequency range $200-25$ MHz). The error bar indicates the variation in the type II height estimate while considering other density models, such as the Baumbach \citep[][two-four fold]{Baumbach1937}, Saito \citep[][two-six fold]{Saito1977} models. This error bar is also similar to the standard deviation in the type II height estimates using a four-fold Newkirk's model.
There were only a few metric type II bursts which started at heights higher than $\approx 2.3 R_{\odot}$ ($\leq 5 \%$) which corresponds to frequencies below 50 MHz. Similarly, there were also fewer type II bursts starting at low corona heights below 1.7~R$_{\odot}$. The majority of type IIs have an onset height between 1.7 and 2.3~R$_{\odot}$ which indicates the height of plasma oscillations due to the electrons accelerated at the magneto-hydrodynamics (MHD) shocks in the solar corona.

       \begin{figure*}
    \centering \includegraphics[width=1\textwidth,clip=]{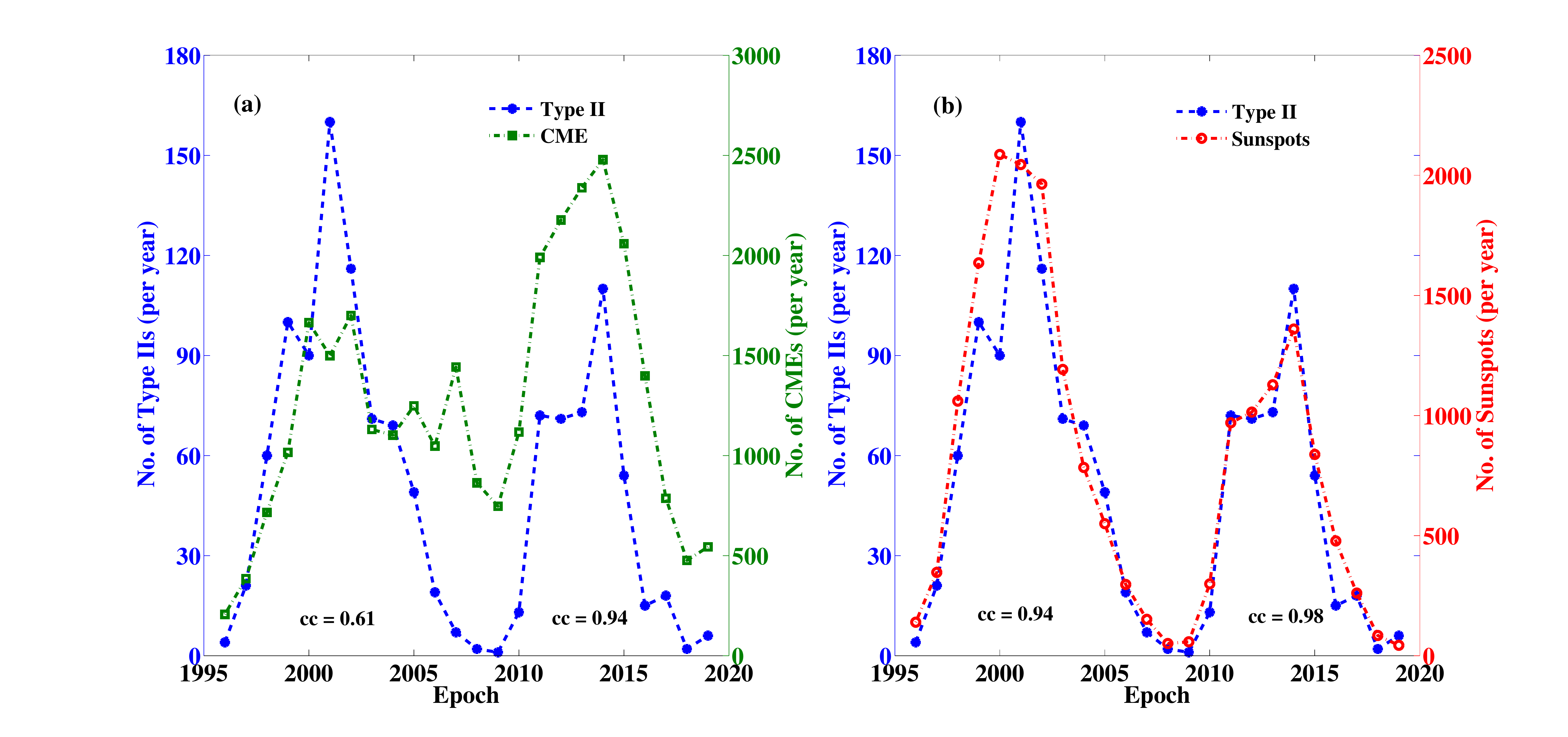}
   \caption{ \textbf{Left panel:} The number of CMEs (per year) and type II bursts (per year) from 1996-2019. The number of type IV bursts are very less as compared to the number of CMEs however, there is a very strong correlation ($94 \%$) between the occurrence of the the white-light feature and radio emission in for cycle 24. There is a marginal correlation between these two quantities in cycle 23. This may be because of lesser CMEs detected by LASCO in the previous cycle. 
    \textbf{Right panel:} The number of type II (per year) and sunspots (per year; adapted from \cite{sidc}) from 1996-2019. These two quantities show a strong correlation in cycle 23 and 24 ($\approx 94 \%$ and $\approx 98 \%$, respectively). }
    \label{Fig:figure7}
    \end{figure*}

\subsection{Type II Bursts, CMEs and Sunspot Number Correlation}
\label{section:sec3.6}

Finally, we studied the relation between CMEs, type II bursts and sunspot number (SSN) during the past two solar cycles. Figure \ref{Fig:figure7}a shows the correlation between number of CMEs and type II bursts recorded per year. The correlation coefficient is low for cycle 23 ($\approx 0.61$), however, type II bursts are highly correlated with CMEs for solar cycle 24 (the coefficient is $\approx 0.95$). 
The sunspot numbers and type II bursts are also highly correlated in both cycles ($\approx 0.94$ and $\approx 0.98$, respectively). Figure \ref{Fig:figure7} also shows that these three phenomenon are quite well correlated for cycle 24. \cite{Michalek2019} had also shown that sunspot number and CMEs were poorly correlated for cycle 23 when compared to cycle 24. The correlation was improved when they separated poor CME events from their list. 

\begin{table*}[]
\centering
\caption{Number and percentage of type II radio bursts associated with CMEs with different properties: 'Fast', 'Slow', 'Wide', 'Narrow', and a combination of these properties.}
\label{tab:table2}
\begin{tabular}{ccccccc}
\hline
Category       & \multicolumn{2}{c}{Total} & \multicolumn{2}{c}{Cycle 23} & \multicolumn{2}{c}{Cycle 24} \\
               & No.         & \%          & No.          & \%            & No.       & \textbf{\%}      \\
               \hline
Total  & 1024        & 100.0\%         & 609           & 59.5\%           & 415       & 40.5\% \\
\hline
Fast           & 554         & 54.1\%         & 347          & 57.0\%           & 207       & 49.9\%             \\
Slow           & 470         & 45.9\%         & 262          & 43.0\%           & 208       & 50.1\%             \\
Wide           & 809         & 79.0\%         & 482          & 79.1\%           & 327       & 78.8\%              \\
Narrow         & 215         & 21.0\%         & 127          & 20.9\%           & 88        & 21.2\%              \\
Fast \& Wide   & 495         & 48.3\%         & 313          & 51.4\%           & 182       & 43.9\%              \\
Fast \& Narrow & 59          & 5.8\%          & 34           & 5.6\%            & 25        & 6.0\%               \\
Slow \& Wide   & 314         & 30.7\%         & 169          & 27.7\%           & 145       & 34.9\%              \\
Slow \& Narrow & 156         & 15.2\%         & 93           & 15.3\%           & 63        & 15.2\% \\
\hline

\end{tabular}
\end{table*}

 \begin{table*}[]
\centering
\caption{Number and percentage of type II radio bursts associated with partial halo and halo CMEs.}
\label{tab:table3}
\begin{tabular}{ccccccc}
\hline
Category          & \multicolumn{2}{c}{Total}                      & \multicolumn{2}{c}{Cycle 23}                   & \multicolumn{2}{c}{Cycle 24}                   \\
                  & No.                  & \%                      & No.                  & \%                      & No.                  & \%                      \\
                  \hline
Total             & 1024                 & 100.0\%                 & 609                  & 59.5\%                  & 415                  & 40.5\%                  \\
\hline
Partial Halo CMEs & 241                  & 23.5\%                  & 131                  & 21.5\%                  & 110                  & 26.5\%                  \\
Halo CMEs         & 252                  & 24.6\%                  & 144                  & 23.6\%                  & 108                  & 26.0\%                  \\
Partial Halo \&   & \multirow{2}{*}{493} & \multirow{2}{*}{48.1\%} & \multirow{2}{*}{275} & \multirow{2}{*}{45.1\%} & \multirow{2}{*}{218} & \multirow{2}{*}{52.5\%} \\
Halo CMEs         &                      &                         &                      &                         &                      &                         \\
Other CMEs        & 531                  & 51.9\%                  & 334                  & 54.9\%                  & 197                  & 47.5\%    \\
\hline
\end{tabular}
\end{table*}

\begin{table*}[]
\centering
\caption{Typical parameters values for type II bursts and CMEs in solar cycle 23 and 24. }
\label{tab:table4}
\begin{tabular}{lccccccccc}
\hline
\multirow{2}{*}{Parameter} & \multicolumn{3}{c}{\multirow{2}{*}{All Type II}} & \multicolumn{3}{c}{\multirow{2}{*}{Type II with CMEs}} & \multicolumn{3}{c}{\multirow{2}{*}{Type II without CMEs}} \\
                           & \multicolumn{3}{c}{}                             & \multicolumn{3}{c}{}                                   & \multicolumn{3}{c}{}                                      \\
                           \hline
                           & Cycle 23        & Cycle 24        & Total        & Cycle 23          & Cycle 24          & Total          & Cycle 23           & Cycle 24           & Total           \\
                           \hline

\multicolumn{5}{l}{Type II Duration (minutes)} \\
Mean                       & 10.4            & 10.7            & 10.5         & 10.7              & 10.8              & 10.7           & 7.5                & 8.4                & 8.2             \\
Median                     & 9.0             & 9.0             & 9.0          & 9.0               & 9.0               & 9.0            & 6.0                & 6.0                & 6.0             \\
Std. Dev.                  & 6.8             & 9.7             & 7.9          & 8.1               & 6.6               & 10.1           & 6.4                & 6.9                & 4.4             \\
\hline
\multicolumn{5}{l}{Type II Drift Rate (MHz/s)} \\
Mean                       & 0.08            & 0.09            & 0.08         & 0.10              & 0.11             & 0.11           & 0.08              & 0.08               & 0.08            \\
Median                     & 0.05            & 0.06           & 0.06         & 0.05              & 0.06              & 0.06           & 0.05             & 0.06               & 0.06            \\
Std. Dev                   & 0.03          & 0.02          & 0.03        & 0.03          & 0.02          & 0.03           & 0.03          & 0.02          & 0.03          \\
\hline
\multicolumn{5}{l}{Type II Speed (km/s)} \\
Mean                       & 700             & 840             & 751          & 645               & 848               & 729            & 632                & 621                & 626            \\
Median                     & 538             & 671             & 592          & 490               & 671              & 563           & 475                & 456               & 461             \\
Std. Dev                   & 568            & 716             & 630          & 773               & 945               & 815            & 773                & 848                & 811            \\
\hline
\multicolumn{5}{l}{Type II Height ($R_{\odot}$)} \\
Mean                       & 1.77            & 1.76            & 1.77         & 1.78              & 1.77              & 1.78           & 1.77
& 1.83               & 1.74            \\
Median                     & 1.71            & 1.62            & 1.69         & 1.73              & 1.64              & 1.69           & 1.69               & 1.85               & 1.67            \\

Std. Dev.                  & 0.27            & 0.29            & 0.28         & 0.26              & 0.29              & 0.28           & 0.27               & 0.27               & 0.28            \\
\hline
\multicolumn{5}{l}{CME Width (Degrees)} \\
Mean                       & --              & --              & --           & 160               & 173               & 165            & --                 & --                 & --              \\
Median                     & --              & --              & --           & 110               & 125               & 115            & --                 & --                 & --              \\
Std. Dev.                  & --              & --              & --           & 126               & 122               & 123            & --                 & --                 & --              \\
\hline
\multicolumn{5}{l}{CME Speed (km/s)} \\
Mean                       & --              & --              & --           & 768               & 670               & 728            & --                 & --                 & --              \\
Median                     & --              & --              & --           & 584               & 499               & 546            & --                 & --                 & --              \\
Std. Dev.                  & --              & --              & --           & 489               & 546               & 526            & --                 & --                 & --    \\
\hline
\end{tabular}
\end{table*}

\section{Discussion}
\label{section:sec4}

We present the first comprehensive long-term analysis of the occurrence of metric type II radio bursts (low coronal radio bursts) during two solar cycles and their association with CMEs of various properties. 
We found that $\approx 79 \%$ and $\approx 95 \%$ of the type II bursts were associated with CMEs in solar cycle 23 and 24, respectively. However,  only $\approx  4 \%$ and $\approx 3 \%$ of CMEs were associated with type II bursts in solar cycle 23 and 24, respectively. 
The results indicate that most of the type II bursts had a white-light CME counterpart, however, there were a few type IIs with no clear CME association. Some studies suggest that such type IIs may be related to flare-driven blast waves \citep{magdalenic2012, pankaj2016, 2023arXiv230511545M}. There were more CMEs in cycle 24 than cycle 23. However, the number of type II radio bursts were less in cycle 24 compared to cycle 23, thus a greater number of CMEs does not indicate that a greater number of type IIs would occur in a solar cycle. The type II occurrence also depends upon the CME properties, such as the lateral expansion, speed, and the coronal conditions such as the ambient magnetic fields (B), Alfv$\acute{\text{e}}$n-Mach number (MA), Alfv$\acute{\text{e}}$n speed ($V_a$), etc. \citep{mann2003formation, lin2006theoretical}. 

Almost $48 \%$ of the type II bursts were associated with 'Fast and Wide' CMEs. This trend is similar to type IV bursts association with CMEs in cycle 24 \citep{Kumari2021} and it was also found in the case of moving radio bursts (mostly type IIs and type IVs) in the study of \citet{Morosan2021}.
There were a very few type II bursts associated with `Slow and Narrow' CMEs. \cite{Morosan2021} suggested that less radio bursts associated with 'Slow and Narrow' CMEs are observed due to the fact that the generation of moving bursts (mostly consisting of type II and type IV bursts) is related to the lateral expansion during the early acceleration phase of a solar eruption. 

Our analysis shows that almost half of the type II bursts in both solar cycles were accompanied with partial and full halo CMEs, essentially they were accompanied by very wide CMEs. Since it is well known that these CMEs can be potentially responsible for geomagnetic storms \citep{Zhang2003}, these type II bursts can be used to study the early kinematics of the geoeffective halo CMEs.
These radio bursts could also be useful in case of stealth CMEs whose early signatures are not visible in coronagraph images, however, so far there has not yet been an account of a type II bursts associated with a stealth CME. The correlation between type IIs and sunspots is also high indicating that type IIs are associated with CMEs resulting from eruptive flares from sunspots and less likely to be associated with phenomena such as filament eruptions that do not require the presence of sunspots \citep{Feynman1994,Choudhary2003}.

In comparison with other statistical studies of radio bursts, there were less type II bursts (435) reported in cycle 24 than type IV bursts \citep[447;][]{Kumari2021}. This could be because for type IV emission, shock is not an essential requirement unlike for type II events \citep{salas2020polarisation}. In turn, a higher number of type II bursts were associated with CMEs compared to type IV bursts. Both type IIs and type IVs are, however, mostly associated with `Fast and Wide' CMEs based on this current and previous studies \citep{Kahler2019,Kumari2021,Morosan2021}. 


Our results suggest that type II bursts dominate at a height of $\approx 1.7 - 2.3 \pm 0.3 ~R_{\odot}$ (between $200-25$ MHz) with a clear majority having an onset height around 1.7 $\pm 0.3~R_{\odot}$ assuming the four-fold Newkirk model. This height range indicates that there may be a critical height for type II occurrence for the majority of solar eruptions. Similar results were found in a theoretical study by \citet{lin2006theoretical} that suggests that type IIs typically occur at a height of 1.5~R$_{\odot}$, which is similar to the heights reported in the present work for the majority of type II bursts. 
The authors modelled a coronal mass ejection (CME) with a terminal speed of approximately 1000 km/s, using magnetic field strength and a constant of magnetic reconnection rate to obtain this height.
\cite{lin2006theoretical} suggested that type II bursts can not appear lower than a critical height, which depends upon then coronal environment and it determines the start frequency of the burst. The critical height is likely highly dependant on the global Alfv\'en speed in the corona. 
For example, at a height of $\sim$1.2 R$_\odot$ near active regions, the Alfv\'en speed is usually greater than 1000 km/s \citep{1988ApJ...330..474P, aschwanden2006physics, morosan2016, regnier2015} and a CME or any other disturbance is too slow to drive a shock at those heights. Below 1.2 R$_\odot$, only five type IIs were found, confirming that CMEs may generally be too slow to become super Alfv\'enic. Even the fastest reported CMEs, require some time to expand in order to reach a speed of 1000 km/s in the low corona \citep[e.g.,][]{mann2003formation, Anshu2017a, Morosan2019a, Pohjolainen2021}. As the CMEs accelerate and expand, they can then reach regions of low Alfv\'en speed \citep[e.g.,][]{Warmuth2004, Morosan2022b} outside the active region associated with the CME eruption. Since type IIs are believed to be the signature of CME-driven shocks, it is likely that CME shocks also form at similar heights. 
The typical type II burst speed was found to be $\approx 750 km/s$ and type II bursts appear to have a higher speed if they have a white-light CME association (see Figure\ref{Fig:figure6}a).
Table \ref{tab:table4} shows the typical parameters such as speed, duration, drift rates for the type II bursts and speed and width for CMEs in SC23 and SC24. The large standard deviation here means that the estimated speeds, drift rates, duration, etc. are widely spread out. Since we have CMEs and type II bursts for all the phases of the solar cycles, it is expected that the deviation in mean parameters will be large. However, it is worth  noting that the median value is relatively close to the mean value \citep{Michalek2019}.

When the CME propagates through the solar corona through the supercritical shock region, the shock wave creates density irregularities in the plasma, which can excite Langmuir waves \citep{Schmidt2012, Mann2022}. Our analysis shows that $\approx 1.7 - 2.3 \pm 0.3 ~R_{\odot}$ (between $200-25$ MHz) represents this supercritical shock region capable of generating electron beams that could excite Langmuir waves. The shock formation heights that generate type II solar radio bursts can vary depending on the properties of CME and the surrounding coronal plasma. To verify independently, we used the Alfv\'en speed, shock height and density relation for type II bursts \citep[for split-band type II radio bursts][]{Smerd1975, 2004A&A...413..753V}. The type II height is directly proportional to start frequency of the burst and magnetic field and inversely proportional to Alfv\'en speed. For an ambient magnetic field of range $1.5 - 2.0$ G \citep[typical B values estimated for type II bursts starting around 200 MHz; see for example][and the references therein]{Dulk1978, kumari2017c}, and Alfv\'en-Mach number in the range $1.59<MA<2.53$ \citep{Mann2022}, we get shock formation height of $\approx 1.7~R_{\odot}$. The lateral expansion during the early acceleration phase of a 'Fast and Wide' CME \citep{Morosan2021, Majumdar2021} can lead to the formation of supercritical shock in the height range $\approx 1.7\pm 0.3 ~R_{\odot}$, which can lead to excitation of Langmuir waves needed for type II solar radio burst emission. For type II bursts, the enhanced Langmuir wave level above the thermal one can be achieved rapidly \citep{Kouloumvakos2021, Mann2022}, compared to type III solar radio bursts which require some travel time for the electrons to reach the region where the Langmuir waves are excited \citep{Reid2020}. This instability distance depends upon the spectral index, electron beam velocity, the size of the acceleration site, and the temporal injection profile. For type II bursts, this happens close to the electron acceleration sites \citep{Schmidt2012, Morosan2022b}.
We note that a few high frequency (heights $\leq 1.5~R_{\odot}$) were observed in the past two solar cycles, but the occurrence of these type II bursts were $< 1 \%$ of the total type II bursts. We note that individual case studies of high frequency type II bursts are required to shred light on this \citep{Mondal2020, Kouloumvakos2021}.



This study extensively uses the reports from SWPC which lists the manual detection of solar radio bursts from the Radio Solar Telescope Network (RSTN) for 24-hour solar observations. However, we acknowledge that RSTN spectrographs have frequency limits and lower sensitivities compared to the new-age non-solar dedicated radio telescopes, such as the LOw Frequency ARray \citep[LOFAR;][]{Haarlem2013} and the Murchison Widefield Array \citep[MWA;][]{Tingay2013}. As our study makes use of two solar cycle radio bursts and CMEs data, we rely on the SWPC, CDAW and SEEDS catalogues for directly extracting the onset time, duration, linear speed, angular width, start and end frequencies, etc. parameters for the events. Since these catalogues are combinations of manual (SWPC and CDAW) and automatic detection (SEEDS) of events, we understand that there may be chances of misclassification of events. We note that we have used linear speed for CMEs and constant drift rate for type II bursts, which may also lead into error in classification of CMEs and type II speeds.
Currently, there are  only two low frequency solar-dedicated imaging instruments, the upgraded Nan{\c c}ay Radioheliograph \citep[NRH;][]{Kerdraon1997} and the Gauribidanur RAdioheliograPH \citep[GRAPH;][]{ramesh1998}, which covers the solar observations between $~ 40-450$ MHz from $2.5-16$ UT at spot frequencies. This still leaves a big temporal window where no solar radio imaging observations are available at present. With the absence of solar-dedicated imaging radio instruments, specially at low radio frequencies, one way to estimate the shock generation locations in the `middle' corona is by using existing density models, such as \cite{newkirk1961solar} (see section \ref{section:sec3.5} for details). This method has been used for in several previous type II radio bursts studies in the absence of imaging observations \citep{vrsnak2004,Vasanth2014,Ramesh2022a}. This study also highlights the importance of having a low radio frequency solar-dedicated imaging instruments for identifying the shock generation locations in the solar corona.


This study should be extended in the future to include imaging observations of type II radio bursts. With dedicated solar radio imaging instruments like the upgraded NRH, the GRAPH, the expanded Owens Valley
Solar Array \citep[EOVSA;][]{Gary2018},
the Mingantu Spectral Radio Heliograph \citep[MUSER;][]{Yan2021} and the upcoming Daocheng Solar Radio Telescope \citep[DSRT;][]{Yan2022cosp}, and other powerful non solar radio imaging instruments like the LOFAR, the MWA, the Long Wavelength Array \citep[LWA;][]{Ellingson2009} and the upgraded Giant Metrewave Radio Telescope \citep[GMRT;][]{Swarup1991,Gupta2017},
multiple datasets can be combined to determine the origin and source regions of these radio bursts with 
better accuracy. With continuous spectroscopic imaging, the source locations of these bursts relative to the CME and their speed relative to the CME flanks can also be investigated in detail to further investigate the formation heights and typical speeds of type II bursts that were also identified in this study using simple empirical models. 

\begin{acknowledgements}
{A.K. and D.E.M. acknowledge the University of Helsinki Three Year Grant. D.E.M acknowledges the Academy of Finland project `RadioCME' (grant number 333859). E.K.J.K. and A.K. acknowledge the European Research Council (ERC) under the European Union's Horizon 2020 Research and Innovation Programme Project SolMAG 724391. E.K.J.K acknowledges the Academy of Finland Project SMASH 310445. All authors acknowledge the Finnish Centre of Excellence in Research of Sustainable Space (Academy of Finland grant number 312390). A.K’s research was supported by an appointment to the NASA Postdoctoral Program at the the NASA Goddard Space Flight Center (GSFC).   }
\end{acknowledgements}


\end{document}